\documentclass{article}
\usepackage{graphicx}
\usepackage{multicol}
\usepackage{amsmath,amssymb,amsfonts}
\usepackage{amsthm}
\usepackage{mathrsfs}
\usepackage[title]{appendix}
\usepackage{xcolor}
\usepackage{textcomp}
\usepackage{manyfoot}
\usepackage{booktabs}
\usepackage{algorithm}
\usepackage{algorithmicx}
\usepackage{algpseudocode}
\usepackage{listings}
\usepackage[font=footnotesize]{caption}
\usepackage[utf8]{inputenc}
\usepackage[T1]{fontenc}
\usepackage[a4paper, margin=1.5cm]{geometry}

\begin{document}
\date{}
\title{Engineering nanowire quantum dots with iontronics.}
\maketitle

\begin{center}
    Domenic Prete$^{1*}$, Valeria Demontis$^1$, Valentina Zannier$^1$, Lucia Sorba$^1$, Fabio Beltram$^1$, and Francesco Rossella$^{1,2}$
    \newline
    \newline
    $^1$ NEST, Scuola Normale Superiore and Istituto Nanoscienze-CNR, Piazza San Silvestro 12, I-56127, Pisa, Italy\\
    $^2$ Scuola di Ingegneria - Dipartimento di Scienze Fisiche, Informatiche e Matematiche, Università di Modena e Reggio Emilia, via Campi 213/a, 41125 Modena, Italy\\
    $^*$ Present address: Scuola di Ingegneria - Dipartimento di Scienze Fisiche, Informatiche e Matematiche, Università di Modena e Reggio Emilia, via Campi 213/a, 41125 Modena, Italy
\end{center}

\textbf{Achieving stable, high-quality quantum dots has proven challenging within device architectures rooted in conventional solid-state device fabrication paradigms. In fact, these are grappled with complex protocols in order to balance ease of realization, scalability, and quantum transport properties.
Here, we demonstrate a novel paradigm of semiconductor quantum dot engineering by exploiting ion gating. Our approach is found to enable the realization and control of a novel quantum dot system: the iontronic quantum dot. Clear Coulomb blockade peaks and their dependence on an externally applied magnetic field are reported, together with the impact of device architecture and confinement potential on quantum dot quality. Devices incorporating two identical quantum dots in series are realized, addressing the reproducibility of the developed approach. The iontronic quantum dot represents a novel class of zero-dimensional quantum devices engineered to overcome the need for thin dielectric layers, facilitating single-step device fabrication. Overall, the reported approach holds the potential to revolutionize the development of functional quantum materials and devices, driving rapid progress in solid state quantum technologies.}
    
\begin{multicols}{2} 
    Quantum technologies have emerged as a critical enabling technology of our era and a top priority for both research and industry, potentially offering unprecedented advancements that will significantly impact society~\cite{Heinrich2021, Raino2021}. Within this context, the development and demonstration of better-performing quantum platforms, which enable the study of physical phenomena at their most fundamental quantum level, are crucial to unlocking these extraordinary opportunities~\cite{Laucht2021}. To accelerate progress in quantum technologies, it is crucial to leverage the extensive expertise gained in nano- and micro-electronics using semiconducting materials: decades of development in manipulating semiconductor-based devices have led to a growing interest in exploring the quantum realm of semiconductors. Quantum dots, specifically, are highly relevant in this context as they are used to implement Single Electron Transistors (SETs) and serve as the fundamental building blocks for semiconductor-based quantum computing~\cite{Pribiag2013, Xue2019, Angus2007, Rossella2014}. Among the various techniques available for implementing semiconductor quantum dots, hard-wall quantum dots in heterostructured nanowires and electrostatically defined quantum dots in 2DEGs and nanowires, in both CMOS compatible silicon architectures~\cite{Zalba2021} and III-V semiconductors~\cite{Chatterjee2021, Burkard2021, Prete2019a, Dorsch2020}, are particularly significant.
    
    However, semiconductor-based quantum technologies are struggling to gain momentum compared to counterparts based on other materials~\cite{DeMichielis2023}. Despite the substantial progress achieved in the development of high-quality semiconductors, device fabrication has largely leaned towards conventional techniques based on the standard MOSFET architecture. This is noteworthy considering the recent advancements in device nanoelectronics, which have capitalized on a novel approach known as iontronics~\cite{Bisri2017, Liang2022, Wang2021}. In the field of iontronics, strong electric fields are applied conformally at the interface between a semiconductor and an electrolyte --- such as an ionic liquid, polymeric electrolyte, or ceramic electrolyte --- achieved by controlling the spatial distribution of ions in the electrolyte and inducing ionic accumulation and thus strong electric fields at the semiconductor/electrolyte interface~\cite{Fujimoto2013, Du2015}. Iontronic devices offer several advantages, namely i) maximized gate capacitance due to the direct contact between the electrolyte and the semiconductor, ii) high robustness of electrolytes against dielectric breakdown and iii) the absence of electrical noise caused by thermal vibrations when the device temperature is maintained below the freezing temperature of the electrolyte~\cite{Ueno2011, Prete2021, Lieb2019}. These features make ion gating a highly promising technique for providing nanoscale gates in quantum devices and enabling innovative experimental platforms for semiconductor-based quantum technologies~\cite{Mikheev2021, Mikheev2023}.
    
    In this work, we develop and demonstrate the first platform implementing a quantum dot stemming from the combination of nanostructured materials (InAs nanowires) and electrolytes ([Emim][Tf2N] ionic liquid). The unprecedented electric field intensities accessible via ion gating techniques are used to electrostatically define a quantum dot below a metallic finger --- the \textit{confinement finger} --- in dual gated devices. These are realized within a single fabrication step - that is, avoiding multiple steps conventionally needed to realize solid-state devices, in which the quantum dot is electrostatically defined with local metallic finger gates separated from the semiconductor by means of thin oxide layers. We characterize the energetics and confinement features of this novel structure --- the \textit{iontronic quantum dot} --- by observing clear Coulomb blockade features in the magneto-transport measurements and comparing the performance of the dot under several confinement conditions. Our findings report a high quantum dot performance, resulting in the observation of the first Coulomb oscillations with a quality comparable to a hard-wall quantum dot defined in heterostructured nanowires. Furthermore, we probe the feasibility of the approach for the development of more complex structures, by fabricating a device with two quantum dots connected in series by means of a larger semiconductor section and probing the energy spectrum in this case, finding pairs of Coulomb peaks with nearly identical properties evolving in the same way in the magneto-transport measurements. 
    
    Overall our results show that a new way of conceiving quantum nanostructures, combining the technological know-how for semiconductor fabrication and the unprecedented high-performance of novel gating techniques, is possible. The potential applications of the newly developed system are incredibly wide, including novel ways to implement spin qubits and to explore quantum thermodynamics on complex nanoscale systems.
    
    The prototypical device architecture developed in this work is shown in Figure 1. An InAs nanowire - $2\,\mu m $ long, with a diameter of 80 nm and featuring fully classical transport behaviour at low temperature - is contacted with two ohmic contacts for bias voltage $\mathrm{V_{DS}}$ application and current measurements. A short finger --- confinement finger henceforth --- is fabricated between these two contacts in order to locally prevent the field effect caused by the ionic accumulation at the interface between the nanowire and the electrolyte which embeds the entire device --- as depicted in the inset in Figure 1(a). This ionic accumulation is driven by the application of a voltage $\mathrm{V_{IL}}$ to a square counter-electrode fabricated tens of micrometers away from the nanowire. Noticeably, the entire device if fabricated within a single lithographic step. The fabrication substrate consists on a Si++/$\mathrm{SiO_2}$ bilayer which is used as a solid-state backgate, thus allowing to operate the device in a dual-gated configuration\cite{Prete2021}. It is worth mentioning that the two gates (i.e., the ionic gate and the solid-state gate) provide very different gate couplings and that only the backgate is expected to effectively work at low temperatures owing to the relatively high freezing temperature of [Emim][Tf2n] ($\sim$190 K) which prevents the ions due to being frozen in their positions. Instead, the ionic gate is expected to have a very strong influence on the charge carriers in the nanowire tunable at room temperature. Consequently, the device is operated in the so-called \textit{set-and-freeze} regime~\cite{Svensson2014}, meaning that the ionic gate is operated at room temperature and that $\mathrm{V_{IL}}$ is fixed during the device cooldown in order to set a specific charge configuration --- resulting from the specific confinement potentials generated by the ions at the nanostructure surface --- in the nanowire, while the backgate is employed at low temperature in order to probe the charge transport features in the nanostructure and provide an experimental tool to tune the electrochemical potential of charge carriers in the semiconductor. 
    
    In this configuration, by accumulating negative ions at the nanowire/ionic liquid interface, on-demand confinement potential can be realized. As an example, the finite element calculations reported in Figure 1(c) shows the spatial distribution of the electrostatic potential energy corresponding to the confinement potential the electrons in the nanowire undergo. It can be noticed that --- away from the finger --- lower energy areas are defined away from the surface of the nanowire, resulting in elongated channels in which electronic wavefunctions exist. Below the finger, owing to the insulation effect from the strong external electric field applied by the ions, the low energy area acquires an elliptical shape and, most importantly, a potential well is defined. This feature is more evident in the cut (Figure 1(d)) of the above mentioned potential profile, computed in the axial direction of the nanowire. Here, the potential well defined below the confinement finger leads to the formation of a low dimensional system featuring quantized energy levels, indicated by the black dashed lines, showing that discrete states are formed within the potential well and that some levels appear to be degenerate due to the hexagonal cross-section of the nanowire~\cite{Ford2012, Degtyarev2017}. Furthermore, the potential well defined below the finger is separated from the outlying sections of the nanowire by barriers, expected to act as tunnel barriers. This energetic profile, i.e., a 0D system coupled with charge reservoirs by tunnel barriers, is typical for single electron transistors. Indeed, Figure 1(e), showing the dependence of the current flowing in the nanowire $\mathrm{I_{DS}}$ at 4.2 K on $\mathrm{V_{DS}}$ and the backgate voltage $\mathrm{V_{BG}}$, exhibits the typical Coulomb diamond pattern features by SETs.

    
    Figure 2 shows electrical transport measurements of an iontronic quantum dot defined with a confinement finger gate width of 100 nm and a freezing liquid gate voltage $\mathrm{V_{IL,freeze}=-1.5\,V}$. The measurement reported in Figure 2(a) shows 6 clear Coulomb blockade diamonds, the first two being sharper and featuring bigger dimensions compared with the others and providing values for the charging energy ($E_c = 12\pm1$ meV) and quantum level spacing ($\epsilon = 4.5\pm 0.5$ meV) --- reproduced in our numerical calculations --- comparable with the confinement regimes found in hard-wall InAs/InP quantum dots~\cite{Romeo2012}. Interestingly, the first two states feature sharper diamonds when compares to the subsequent ones. We tentatively ascribe this effect to the increase of the backgate voltage having the effect of making the charge carriers wavefunction closer to the bottom facet of the nanowire, where the ionic liquid gating has a weaker effect (since the nanowire is dropcasted on the fabrication substrate). This results in a weaker confinement effect. Consequently, this would cause the states visible after the second being effectively less confined compared to the first two measured states.

    Related to this is that the height of the first two peaks is lower compared to the other Coulomb oscillations, as visible in Figure 2(b), reporting $\mathrm{I_{DS}-V{BG}}$ curves at fixed bias voltages (2, 4 and 6 mV respectively in black, red and blue). We tentatively ascribe this feature to the combination of two phenomena, the first being that the increase of the backgate voltage compensates the depletion effect caused by the ionic accumulation at the nanowire surface, while the second consists in the shift of the quantum dot energy spectrum to higher energy electrochemical potentials. The first effect causes a higher density of states in the leads acting as electron reservoirs for the iontronic quantum dot, corresponding to an increase in the tunneling rates for the source-dot and drain-dot tunneling events. The second effect is responsible of the lower energy separation between the levels after the second, since higher energy electrochemical potentials are found to be degenerate due to the hexagonal cross section of the nanowire. Nonetheless, for practical applications the first few electrons visible as isolated states in the quantum dot are conventionally employed. It is worth mentioning that, despite there is no signal measured for low backgate voltage, we cannot argue that the first Coulomb diamond observed in our system for backgate voltages higher than the threshold voltage $\mathrm{V_{Th}}$ is the very first electron to occupy the iontronic quantum dot, since the backgate has a global impact on both the electrochemical potentials in the quantum dot, as well as on the leads. For this reason, the observation of no current flowing for low $\mathrm{V_{BG}}$ could be related either to the absence of free states in the quantum dot, or to the complete depletion of charge carriers in the semiconducting leads. In order to extract the tunneling rates between the dot and the leads, the conductance of the first two peaks has been fitted with the sequential tunneling lineshape $G = \frac{e^2}{h}\frac{\Gamma}{8k_BT}\frac{1}{cosh^2\left(\alpha (V_{BG}-V_{Th})/2k_BT\right)}$, where $\Gamma$ is the source/drain tunnelling rate (assumed to be equal), $\alpha$ is the backgate lever arm --- extracted from the Coulomb diamond map measurement --- and $k_B$ is the Boltzmann constant. The outcome of the fit is reported in Figure 2(c) and the extracted tunneling rates for the source and drain contacts  are $\Gamma_1 = 2 GHz$ for the first peak and $\Gamma_1 = 5 GHz$ for the second. Interestingly, these values for tunneling rates are coherent with the corresponding values usually reported in literature for hard-wall quantum dots, again confirming the excellent quality of the iontronic quantum dot for the first electrons~\cite{Momtaz2020}.
    
    Similar experimental evidence are found in the magneto-transport measurements, reported in Figures 2(d)-(e)-(f), reporting the dependence of the Coulomb peaks' position on an externally applied magnetic field orthogonal to the NW axis, from 0 T to 8 T. A clear spin up-spin down pattern following the Hund rule is observed, as visible in Figures 2(d)-(e). The shift of the backgate voltage position of the Coulomb blockade peaks is fitted by means of a linear function based on the Zeeman splitting $V_{BG_{pk}}=(V_{BG_{pk}, B=0} \pm g^*\mu _b B)/\alpha$ for spin down and spin up respectively (Figure 2(f)), where $g^*$ is the size-dependent effective g-factor for InAs and $\alpha$ is the backgate lever arm extracted from the Coulomb diamonds size. The same value for $g^*$ is used to fit every peak maximum, and the value $g^*=5.4\pm0.5$ is found --- consistent with InAs quantum dots reported in literature~\cite{Bjork2005}.

    We now address the investigation of the dependence of the quality and sharpness of the observed quantum features on the physical parameters employed to define the system, i.e., the width of the confinement gate $W$ and the freezing liquid gate voltage $\mathrm{V_{IL, freeze}}$. We use finite element analysis to calculate the electric potential in the nanowire with an externally applied electric displacement field generated by the ionic accumulation at the nanostructure's surface, together with the eigenstates of the confined electrons, and the effects on the confinement finger width and intensity of confinement potential on the quantum features of the iontronic quantum dot. Figure 3 reports the main results of such investigation. Specifically, Figures 3(a)-(b) report the normalized electronic wavefunctions of the ground state calculated for several values of the confinement field and a fixed value of the finger width (W = 100 nm) along the axial and the radial directions of the nanowire. Figures 3(c)-(d) report the corresponding results calculated with a fixed value for the confinement field (D = 0.0046 $\mathrm{C/m^2}$) for different values of the confinement finger width. These results show that the confinement field has a much stronger effect on the spatial extension of the electronic wavefunction when compared to the effect of the finger width, suggesting that the applied liquid gate voltage is the main driving force towards the definition of the confined 0D system. In order to probe the combined effect of W and D on the quantum features of the system, the energy difference between the first two states are computed in the parameter space spanned by these quantities, as reported in Figure 3(e). Since the calculations do not take into account any spurious effects, e.g.,  screening or cross-talks between the left/right potential barriers, the simulation trivially shows that the better confinement is achieved by the shortest finger width and the strongest confinement field.
    
    Despite being able to grasp the fundamental features of this novel quantum architecture and to rationalize its behaviour, the numerical results just shown provide a simplified picture of the system, for which experimental evidence shows a more complex scenario --- as reported in the following. In order to reduce as much as possible any variation in the semiconductor properties between measurements, devices featuring more than one confinement fingers separated by ohmic contacts have been fabricated on the same nanowire allowing to probe the iontronic quantum dot properties with different confinement gate widths on the same nanostructure. Devices with confinement finger widths of 60 nm, 100 nm and 125 nm were fabricated and tested by fixing the freezing liquid gate voltage to -1.5 V, -2 V and -2.5 V. All measurements were carried at 4.2 K by testing each iontronic quantum dot featuring different finger sizes before warming up and changing the freezing voltage for the next acquisition. The results are shown in Figures 4(a)-(i), reporting Coulomb blockade maps for each of the investigated configurations in a $\mathrm{\tilde{V}_{DS}-\tilde{V}_{BG}}$ range focusing on multiple Coulomb diamonds. Noticeably, despite Coulomb blockade features are visible for all the data sets, their quality and sharpness are found to be strongly varying with respect to the $W$ and $\mathrm{V_{IL, freeze}}$. Specifically, a trend is noticeable going from the weaker confinement regime (-1.5 V) to more negative freezing liquid gate voltages, showing that in the latter case the quality of the quantum structure is decreased, with a more pronounced effect in the case of wider the confinement finger. Furthermore, the best iontronic quantum dot was observed in the case $W=100\,nm$ and $\mathrm{V_{IL, freeze}=-1.5\,V}$.
    In order to quantify this observation, each Coulomb blockade map was analyzed by extracting a figure of merit parameter Q, computed as the product of the first diamond height with the total number of observable Coulomb diamonds before reaching the continuum states of the spectra, normalized to 1. The results have been interpolated to produce the map reported in Figure 4(j). Overall, the experimental results are coherent with their numerical counterparts on the stronger effect of the confinement field when compared to the finger width. However, it is possible to appreciate a hot-spot in the (100 nm, -1.5 V) area of the plot, corresponding to the best iontronic quantum dot measured in the parameter space spanned by W and $\mathrm{V_{IL, freeze}}$. By increasing the confinement potential (i.e., moving towards more negative liquid gate voltages) the quality of the iontronic quantum dot is reduced both for W=125 nm and W=100 nm, while it is slightly improved for W=60 nm. This behavior can be tentatively ascribed to the combination of the two effects discussed previously, which are not taken into consideration in the numerical calculations: the depletion of the leads acting as reservoirs for the quantum dot and the confinement profile of the quantum dot itself. Indeed, for more negative liquid gate potentials, the leads are much more depleted compared to the configurations in which the system is frozen with a more positive applied liquid gate voltage. Since a stronger back-gate voltage is needed to compensate the depletion induced by the ionic liquid gate, the electrical conduction through the iontronic quantum dot is enabled by higher energetic and less separated states, due to the initial and better separated electrochemical potential levels have been shifted by the back-gate voltage. Additionally, when smaller confinement finger widths are concerned, a certain amount of inter-barrier cross-talk can be envisioned, leading to detrimental effects on the quantum confinement of charge carriers. However, for both practical applications and for qualitatively better results, it is better to resort to a configuration in which small back-gate voltages are needed to observe quantum transport in the iontronic quantum dot (i.e., $\mathrm{V_{IL, freeze}=-1.5\,V}$) and the confinement finger width is long enough to avoid any cross-talk between the tunnel barriers and short enough to enable the potential well defining the quantum dot to be deep and observe well-spaced levels, as in the case for W = 100 nm.

    The iontronic quantum dot provides a highly flexible and easily scalable approach to fabricate quantum devices based on single semiconducting nanostructures without having to rely on complex series fabrication steps. To demonstrate so, a device fabricated with a single InAs nanowire featuring a series to two iontronic quantum dots was fabricated, as shown in Figure 5(a). Here, two 100 nm wide confinement fingers are fabricated on the nanowire such that the distance between them is long enough (1 $\mathrm{\mu m}$) to prevent any coherent transport or tunneling between the two quantum systems. The rationale of this device is to measure the electronic transport through a system as represented in Figure 5(b). Since the two quantum dots are decoupled from each other, one would expect to measure two identical quantum dots in series, i.e., electrical transport is enabled when the electrochemical potentials of the two quantum dots - e.g., $\mu_{n,LQD}$ and $\mu_{n,RQD}$ in Figure 5(b) - are aligned. Since the two iontronic quantum dots are nominally identical (the two confinement finger are fabricated in a single fabrication step together with the ohmic contacts and the same droplet of ionic liquid is covering both structures) the pattern observed in the electrical transport for the quantum features is expected to be the same as for a single iontronic quantum dot. Indeed, panels in Figure 5(c)-(e), reporting bias spectroscopy measurement of the device under investigation at 4.2 K with a freezing liquid gate voltage $\mathrm{V_{IL, freeze}=-1.5\,V}$, show a series of nearly identical pairs of Coulomb diamonds in the Coulomb blockade map (Figure 5(c)), featuring properties coherent with the single iontronic quantum dot measurements presented in Figure 2. The non-ideal overlap of the electrochemical energy spectra of the two dots is ascribable in differences in the local electrostatic landscape in the nanowire in the two points where the quantum dots are formed, which may be caused by inhomogeneities in the surface states by impurities in the semiconductor crystal lattice, causing a small rigid shift between the energy spectra of the two quantum dots. Figure 5(d) allows to observe that these paired Coulomb peaks overlap and a fit of the tunneling rates for the first two peaks (as highlighted in Figure 5(e)) returns $\Gamma_1 = 5.4 \pm 0.5 GHz$ and $\Gamma_2 = 5.7 \pm 0.5 GHz$ for the first and second peak, respectively. These values are consistent with each other, suggesting that they are coming from corresponding levels in the two quantum dots. The picture presented here is further confirmed by the magneto-transport measurements shown in Figure 5(f), where the paired peaks evolve in a parallel fashion in an externally applied magnetic field (from 0 T to 8 T), reflecting that they refer to parallel spins isolated in the two iontronic quantum dots. 

    \section*{Conclusions}
    To conclude, in this work a novel class of quantum device architecture has been demonstrated and characterized, i.e. the iontronic quantum dot. This device concept exploits the unprecedented strong electric fields accessible with ion-gating to define high quality 0D systems in an otherwise classical semiconductor nanostructure with ease, without resorting to complex multi-step fabrication processes. The quantum confinement is provided by accumulating negative ions at the electrolyte/nanowire interface, except for a small segment of the semiconductor where a thin finger is fabricated in order to prevent the shift of the Fermi energy. The system is cooled down in this configuration, and at low temperature the ionic configuration is fixed in place due to the physical properties of the electrolyte, allowing to ensure high stability for the confinement potentials. In this way the novel structure of the iontronic quantum dot is realized. Bias spectroscopy and magneto-transport measurements have been performed on this new system, finding that the quality and stability of the confinement enabled by ion gating can go on par with hard-wall quantum dots, in an homogeneous material. To demonstrate the scalability of the approach, a device implementing two iontronic quantum dots in series has been realized and characterized, showing that a virtually arbitrary number of high quality quantum dots can be realized on a single crystal semiconductor with ease. 
    This work introduces the realization of an entire new class of quantum devices exploiting the favorable properties of ion gating to implement tailored quantum wavefunctions for charge carriers in semiconductors, in which the charge densities positions are precisely controlled thanks to the strong electrostatic interaction with the ions, providing the unprecedented field effect and allowing to accumulate charge carriers far from the surface of the semiconductor, where surface states and impurities hamper their electrical transport properties. The developed platform features a high level of engineerability, allowing for the optimization of ideally every component, being material-agnostic and generally applicable on every semiconductor. Furthermore, the possibility to employ polymeric electrolytes instead of ionic liquids would enable to address the quantum confinement of single quantum systems independently.
    This work pushes forward the frontiers of quantum technologies, enabling the realization of on-demand high-performance quantum systems featuring ease of fabrication and superior electrical transport properties, with potential applications in quantum computation and quantum information platforms.
      	
  	\bibliographystyle{naturemag}
	\bibliography{References}
\end{multicols}

\newpage

\begin{figure}[b!]
\centering
\includegraphics[width=\textwidth]{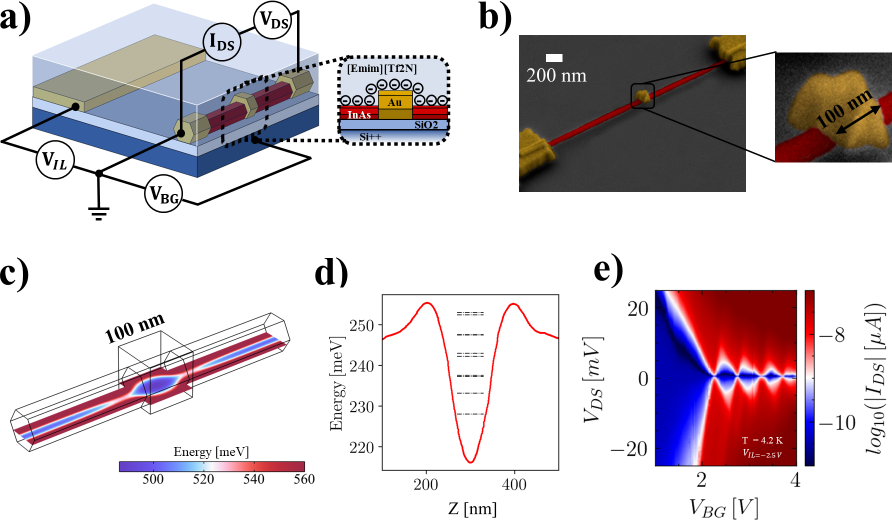}
\caption{Structure, working principle and electrical transport measurement of a prototypical iontronic quantum dot device developed in this work. (a) A pictorial view, measurement circuital schematic, and cross-section of the device, which is based on a dual-gated structure that combines a conventional back-gate and an ionic liquid-based gate. The entire device is immersed in a droplet of ionic liquid, and the ionic distribution is managed via the application of a voltage to a counter-electrode that is co-planar with the nanowire. (b) Tilted scanning electron micrograph of a prototypical device is shown (inset: zoom of the screening finger covering the entire lateral surface of the nanowire). (c) Numerical calculation of spatial energy distribution in a longitudinal cross-section of the nanowire. A low-energy area is formed below the confinement finger. (d) Energy profile along the axial direction of the nanowire extracted from the distribution reported in (c). Dashed lines correspond to the calculated discrete energy levels of the confined single particle states enabling the quantum dot transport features. (e) Coulomb blockaded current measurement in the iontronic quantum dot (T = 4.2 K, $\mathrm{V_{IL}, freeze} = -2.5\, V$, W = 100 nm).}
\label{fig1}
\end{figure}
\newpage
\begin{figure}[tb!]
\centering
\includegraphics[width=0.8\textwidth]{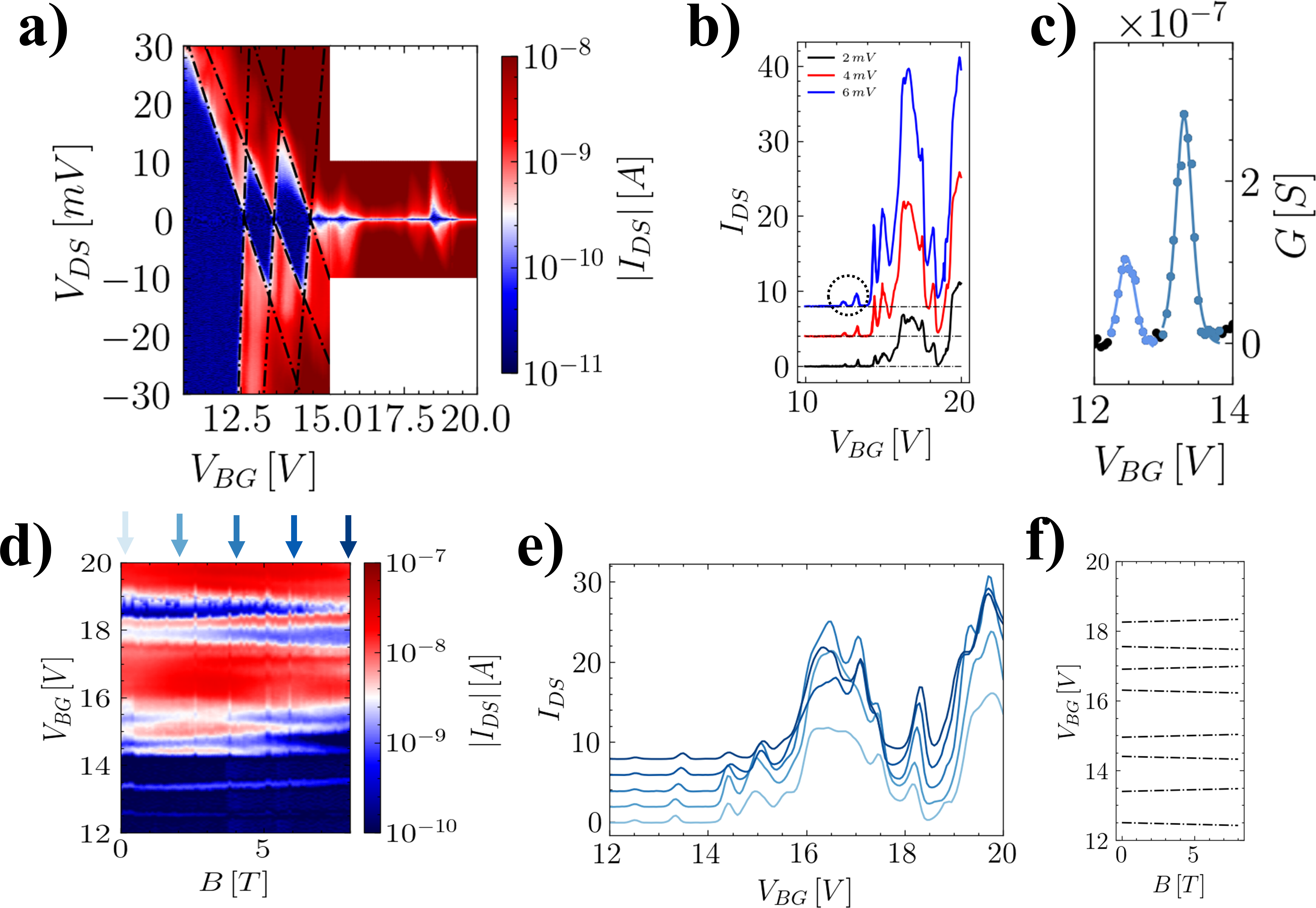}
\caption{Iontronic quantum dot transport spectroscopy and magneto-spectroscopy ( T= 4.2 K,  $\mathrm{V_{IL}, freeze} = -1.5\, V$ ,W = 100 nm). (a) Coulomb diamonds at 0 applied magnetic field. Up to 6 Coulomb diamonds are visible, and from the first two diamonds the charging energy and level spacing of 12.0 meV and 4.5 meV are extracted. (b) Coulomb blockade peaks measured with a bias voltage of 2 mV (black), 4 mV (red) and 6 mV (blue). (c) Best fit on the first two Coulomb peaks measured with an applied bias of 6 mV. From the fit the values for the tunneling rates of the two electrochemical levels are extracted. (d) Electrical transport measurement at fixed $\mathrm{V_{DS}=2\,mV}$ and in an externally applied  magnetic field spanning from 0 to 8 T. (e) Line cuts extracted from panel (d) at different values of magnetic field according to the arrows colour code. (f) Best fit of the linear dependence of the Coulomb peaks with the applied magnetic field, showing that level filling is consistent with the Hund rule.}
\label{fig2}
\end{figure}
\newpage
\begin{figure}[tb!]
\centering
\includegraphics[width=0.8\textwidth]{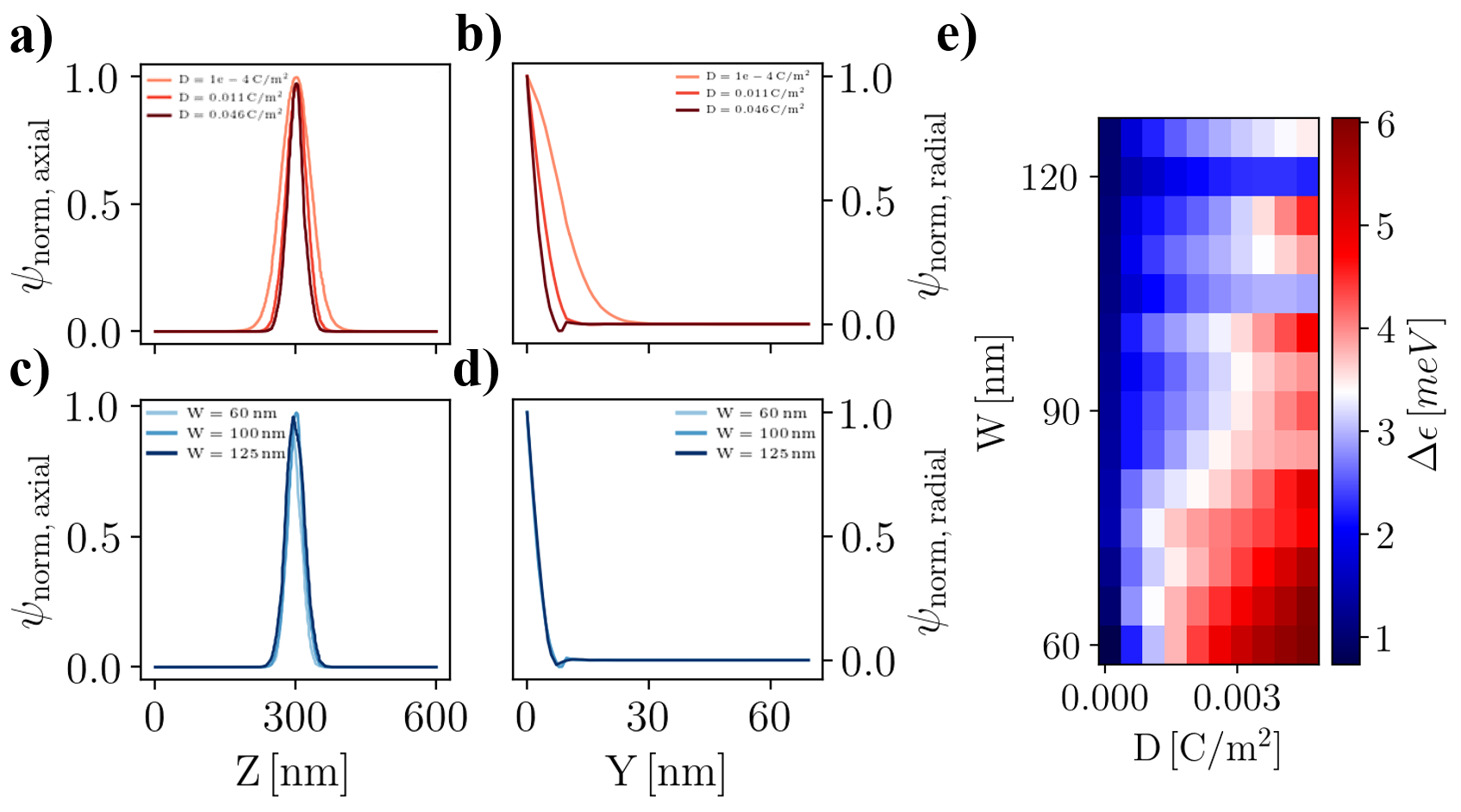}
\caption{Spatial profiles of the Iontronic quantum dot ground state wavefunction, calculated by varying the confinement finger width W and confinement field value D. a-b) Normalized ground state wavefunction lineshape along the axial  (a) and radial (b) direction of the nanowire calculated for different values for the confinement field, showing that this parameter has a strong impact on the spatial extension of the quantum state. c-d) Corresponding calculations of the ground state wavefunctions for several values of confinement finger widths, showing that the effect on the wavefunction's spatial extension is strongly suppressed. e) Quantum level spacing between the first two confined levels in the iontronic quantum dot $\Delta \epsilon$ as a function of W and D.}
\label{fig2}
\end{figure}
\newpage
\begin{figure}[tb!]
\centering
\includegraphics[width=0.8\textwidth]{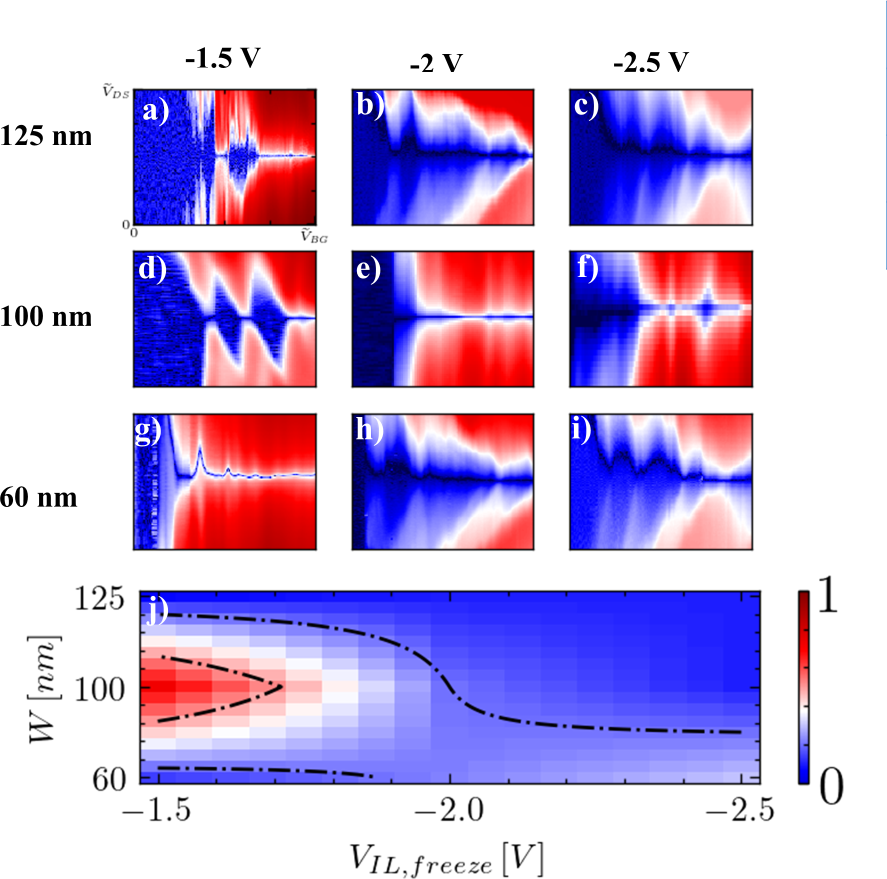}
\caption{Experimental evolution of the iontronic quantum dot stability diagram and figure of merit Q with the confinement finger width W and confinement voltage $\mathrm{V_{{IL},freeze}}$. (a)-(i) Coulomb blockade maps for 3 different values of the confinement finger width (125 nm, 100 nm, 60 nm) and confinement voltage (-1.5 V, -2 V and -2.5 V). (j) Iontronic quantum dot quality factor in the parameter space spanned by confinement finger width and confinement voltage.}
\label{fig3}
\end{figure}
\newpage
\begin{figure}[tb!]
\centering
\includegraphics[width=0.8\textwidth]{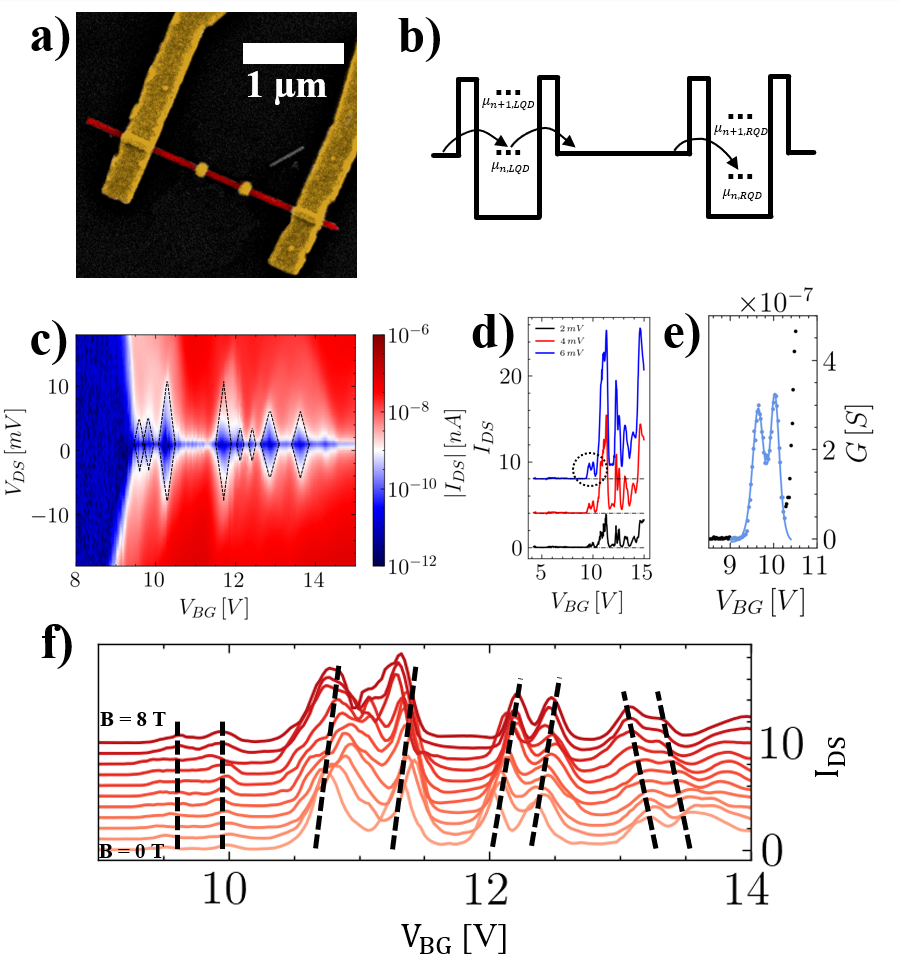}
\caption{Electrical transport and magneto-transport measurements of two identical iontronic quantum dots fabricated in series. (a) Scanning electron micrograph of a prototypical device: two 100 nm wide confinement fingers 1 $\mu m$ away from each other are fabricated on the same nanowire. (b) Pictorial energy diagram of the series quantum dots at 0 applied bias voltage, evidencing identical energy spectra. (c) Bias spectroscopy measurement revealing pairs of Coulomb diamonds with comparable size. (d) Coulomb peaks measured for $V_{DS}$=2 mV (black), 4 mV (red) and 6 mV (blue). Paired peaks with the same height are observed. (e) Best fit of the first two Coulomb peaks measured for $V_{DS}$=6 mV. (f) Coulomb peaks measured with an externally applied magnetic field from 0 to 8 T: paired peaks evolve in a parallel fashion as indicated by the black dashed lines.}
\label{fig4}
\end{figure}

\end{document}